# Reworking The Antonsen-Bormann Idea

S.G.Kamath*

Department of Mathematics, Indian Institute of Technology Madras

Chennai 600 036, India

*e-mail:  kamath@iitm.ac.in

Abstract:   The Antonsen – Bormann idea was originally proposed by these authors for the computation of the heat kernel in curved space; it was also used by the author  recently with the same objective but for the Lagrangian density for a real massive scalar field in 2 + 1 dimensional curved space. It is now reworked here with a different purpose – namely, to determine the zeta function for the said model using the Schwinger operator expansion.

## Introduction

This is the third of a three part paper [1, 2] dealing with the Lagrangian density

$$L = \tfrac{1}{2} g^{\mu\rho} \partial_\mu \varphi \partial_\rho \varphi - \tfrac{1}{2} m^2 \varphi^2 \tag{1}$$

for a real massive scalar field in 2 + 1 dimensional curved space. To motivate this paper we consider the operator

$$B \equiv -\partial^\mu (g_{\mu\rho} \partial^\rho) - m^2 \tag{2}$$

and work with the stationary solutions of the Einstein field equations given by Deser et al.[3] and Clement[4] to define the metric  $g^{\mu\nu}$; in detail

$$g^{00} = 1 - \tfrac{\lambda^2}{r^2}, g^{01} = -\tfrac{\lambda y}{r^2}, g^{02} = \tfrac{\lambda x}{r^2}, g^{11} = -1, g^{12} = 0, g^{22} = -1 \tag{3a}$$

and  $g_{\mu\nu}$  is

$$g_{00} = 1, g_{01} = -\tfrac{\lambda y}{r^2}, g_{02} = \tfrac{\lambda x}{r^2}, g_{11} = -1 + (\tfrac{\lambda y}{r^2})^2 \quad g_{12} = -\left(\tfrac{\lambda^2 xy}{r^4}\right), g_{22} = -1 + (\tfrac{\lambda x}{r^2})^2 \tag{3b}$$

with  $r = |\vec{r}|$  , $2\pi\lambda = \kappa J$, and   $\kappa = 8\pi G$ ; G  being the gravitational constant and $J = |\boldsymbol{J}|$ being the spin of the massless particle located at the origin(see eq.(20) in  Clement[4]).

The effort in Ref.2 was to the use of the Antonsen-Bormann idea [5,6] for the heat kernel $G(x, x'; \sigma)$ in planar curved space , it being the solution of

$$BG(x, x'; \sigma) = -i \tfrac{\partial G}{\partial \sigma} \tag{4}$$

with $G(x, x'; \sigma \to 0) = -i \delta^{(3)}(x - x')$ ; by introducing the vierbeins $e_a^\alpha$ defined by $g^{\alpha\beta} = \eta^{ab} e_a^\alpha e_b^\beta$ with $\eta^{ab} = \text{diag}(1, -1, -1)$ one first reworks B as

$$B = -\eta^{ab}\partial_a\partial_b - m^2 - e_\alpha^m \partial_m(e_a^\alpha)\partial^a \tag{5}$$

Writing $G(x, x'; \sigma)$ in terms of the flat space heat kernel $G_0(x, x'; \sigma)$ as

$$G(x, x'; \sigma) = G_0 e^{-\frac{1}{2}\int e_\alpha^m \partial_m(e_n^\alpha) dx^n} e^{-T} \tag{6}$$

with $(\eta^{ab}\partial_a\partial_b + m^2)G_0 = i\frac{\partial G_0}{\partial \sigma}$ and $G_0(x, x'; \sigma \to 0) = -i\delta^{(3)}(x-x')$, one then obtains as in Ref.5

$$\partial^a\partial_a T - \partial^a T \partial_a T + f + 2\frac{\partial^\mu G_0}{G_0}\partial_\mu T = i\frac{\partial T}{\partial \sigma} \tag{7}$$

with $f = \frac{1}{4}(e_\alpha^m \partial_m(e_b^\alpha))^2 + \frac{1}{2}\partial^n(e_\beta^m \partial_m(e_n^\beta))$. Since $G_0 = (4\pi i\sigma)^{-3/2} e^{-i\frac{(x-x')^2}{4\sigma} - im^2\sigma}$ the ratio in (7) works to $-i\frac{(x-x')^\mu}{\sigma}\partial_\mu T$. With T taken as

$$T = \frac{\tau_{-1}}{\sigma} + \sum_{k=0}^{\infty} \tau_k \sigma^k$$

and not as in eq.(24) in Ref.5 to meet the twin requirements of the boundary condition on the heat kernels G and $G_0$ as $\sigma \to 0$ thus setting $\tau_0 = -\frac{1}{2}\int e_\alpha^n \partial_n(e_m^\alpha) dx^m$ and retaining the extra term $-i\frac{(x-x')^a}{\sigma}\partial_a T$ one gets $\tau_{-1}$ as a solution of

$$\partial^a\partial_a \tau_{-1} - 2\partial^a \tau_0 \partial_a \tau_{-1} - i(x - x')^a \partial_a \tau_0 = 0 \tag{8}$$

with the $\tau_k$ for $k \geq 1$ being likewise obtained[2] through the solution of similar coupled partial differential equations. This latter aspect is clearly a disincentive to the above program of obtaining the heat kernel $G(x, x'; \sigma)$ ; but we believe the idea [5,6] itself is too good to ignore and it is therefore appreciated below from a different point of view.

**The reworking**

To elaborate, let's return to the operator B as given in eq.(5) and write it as

$$B = H_0 + H_I, H_0 \equiv -\eta^{\mu\nu}\partial_\mu\partial_\nu - m^2, \ H_I \equiv -e_\alpha^m \partial_m(e_n^\alpha)\partial^n \tag{9}$$

We now present two sets of vierbeins $e_a^\mu$ and $e_\mu^a$ that are easily obtained from eqs.(3) above using $g^{\alpha\beta} = \eta^{ab} e_a^\alpha e_b^\beta$ and $g_{\alpha\beta} = \eta_{ab} e_\alpha^a e_\beta^b$, together with the respective operator $H_I$ for each set :

**Set 1:** $e_a^\mu$:

$$e_0^0 = 1, e_1^0 = -\frac{\lambda}{r}, e_2^0 = 0 \qquad e_0^0 = 0, e_1^0 = i, e_2^0 = 0$$

$$e_0^1 = 0, e_1^1 = -\frac{y}{r}, e_2^1 = \frac{x}{r} \qquad e_\mu^a : \quad e_0^1 = -i, e_1^1 = i\frac{\lambda y}{r^2}, e_2^1 = -i\frac{\lambda x}{r^2}$$

$$e_0^2 = 0, e_1^2 = \frac{x}{r}, e_2^2 = \frac{y}{r} \qquad e_0^2 = 0, e_1^2 = 0, e_2^2 = 1$$

$$H_I = -\frac{1}{r^3}\{x(\lambda + iy)p_1 + (\lambda y - ix^2)p_2\} \tag{10a}$$

**Set 2:** $e_a^\mu$:

$$e_0^0 = 1, e_1^0 = \frac{\lambda x}{r^2}, e_2^0 = \frac{\lambda y}{r^2} \qquad e_0^0 = 0, e_1^0 = -\frac{i}{\sqrt{2}}, e_2^0 = \frac{i}{\sqrt{2}}$$

$$e_0^1 = 0, e_1^1 = 0, e_2^1 = 1 \qquad e_\mu^a : \quad e_0^1 = -i, e_1^1 = \frac{i\lambda y}{r^2}, e_2^1 = -\frac{i\lambda x}{r^2}$$

$$e_0^2 = 0, e_1^2 = -1, e_2^2 = 0 \qquad e_0^2 = 0, e_1^2 = -\frac{1}{\sqrt{2}}, e_2^2 = -\frac{1}{\sqrt{2}}$$

$$H_I = -\frac{\lambda}{r^4}\{(y^2 - x^2)p_1 - 2xyp_2\} \tag{10b}$$

with $p_i = -i\partial_i$, i= 1,2 in eqs.(10). Note that eq.(10a) is linear in $\lambda$ while the operator $H_I$ in (10b) is a multiple of $\lambda$; this difference between the two operators in eqs.(10) will be decisive in the sequel.

### The Schwinger Expansion

To take up the subject of this paper, we adopt the method of operator regularization – a perturbative expansion introduced by Schwinger[7] to compute amplitudes in quantum field theory in the context of background field quantization. Specifically, we shall use the method of operator regularization [8,9] to attempt a calculation of the $\zeta(s)$ function associated with the Lagrangian density in eq.(1). To this end we recall the definition of the $\zeta$ function for the operator B following McKeon and Sherry [8]

$$\zeta(s) = \frac{1}{\Gamma(s)} \int_0^\infty dt \ t^{s-1} \ tr \ e^{-Bt} \tag{11}$$

with the functional trace being computed in momentum space after Schwinger[7]. For convenience we shall work in Euclidean space below and rewrite eq.(9) as $B = C_0 + C_1$, with

$$C_0 = p_0^2 - m^2 \quad , \quad C_1 = \vec{p^2} + H_I \tag{12}$$

Since the metric $g^{\mu\nu}$ in eqs.(3) is time independent it is obvious that

$$e^{-(C_0 + C_1)t} = e^{-C_0 t} e^{-C_1 t} \tag{13}$$

The Schwinger expansion [8] is now applied to the second exponential in (13) to get,

$$e^{-t\,C_1} = e^{-t\,p^2} + (-t)\int_0^1 d u\, e^{-t\,(1-u)p^2} H_I\, e^{-t\,u\,p^2} + (-t)^2 \int_0^1 u\, du \int_0^1 du_1 e^{-t\,(1-u)p^2} H_I e^{-t\,u(1-u_1)p^2} H_I\, e^{-t\,u\,u_1 p^2}$$

$$+ (-t)^3 \int_0^1 u^2 du \int_0^1 u_1 d u_1 \int_0^1 du_2\, e^{-t\,(1-u)p^2} H_I e^{-t\,u(1-u_1)p^2} H_I e^{-t\,u u_1(1-u_2)p^2} H_I e^{-t\,u u_1 u_2 p^2}$$

$$+ (-t)^4 \int_0^1 u^3 du \int_0^1 u_1^2 du_1 \int_0^1 u_2 d u_2 \int_0^1 du_3\, e^{-t\,(1-u)p^2} H_I e^{-t\,u(1-u_1)p^2} H_I e^{-t\,u u_1(1-u_2)p^2} H_I$$

$$\times e^{-t\,u u_1 u_2(1-u_3)p^2} H_I e^{-t\,u u_1 u_2 u_3 p^2} + \dots \tag{14}$$

The scope of this paper is limited to the third order term in $(-t)$ above with the higher order terms being dealt with in more detail elsewhere. An obvious advantage from the use of (10b) over (10a) in (14) is that it becomes a power series in $\lambda$ and this favours the use of eq.(10b) for the calculation done below.

Writing the first order term in (14) as

$$(-t) \int_0^1 du\, e^{-t(1-u)p^2} \langle p|H_I|p\rangle e^{-t\,u\,p^2} \tag{15}$$

The matrix element in (15) now becomes,

$$\langle p|H_I|p\rangle = -\lambda \int \frac{1}{(q_1^2+q_2^2)^2}((q_2^2-q_1^2)p_1 - 2q_1 q_2 p_2) \tag{16}$$

with $q_1$ and $q_2$ being the two components of the position vector $\vec{q}$ and $\int$ being a symbol for $\int \frac{d^2 q}{(2\pi)^2}$. On integration the first and second terms in (16) yield zero by symmetry ; thus the first order term in (14) is zero. The second order term in (14) is given by

$$(-t)^2 \int_0^1 u\, du \int_0^1 du_1 e^{-t\,(1-u)p^2} \int \langle p|H_I|r\rangle e^{-t\,u(1-u_1)r^2} \langle r|H_I|p\rangle e^{-t\,u u_1 p^2} \tag{17}$$

with $\vec{r}$ being a two component momentum vector and $\int$ shorthand for $\int d^2 r$. As in (16), one gets

$$\langle p|H_I|r\rangle = -\lambda \int \frac{e^{-i(\vec{p}-\vec{r})\cdot\vec{q}}}{(q_1^2+q_2^2)^2}\{(q_2^2-q_1^2)r_1 - 2q_1 q_2 r_2\} = -\frac{\lambda}{4\pi}\left\{r_1 - 2\frac{(p_2-r_2)(p_2 r_1 - p_1 r_2)}{(\vec{p}-\vec{r})^2}\right\} \tag{18}$$

Therefore

$$\langle r|H_I|p\rangle = -\frac{\lambda}{4\pi}\left\{p_1 - 2\frac{(r_2-p_2)(r_2 p_1 - r_1 p_2)}{(\vec{r}-\vec{p})^2}\right\} \tag{19}$$

With eqs. (18) and (19) , (17) is easily determined from the three apparently nonzero momentum integrals:

$$J_0 \equiv (-2)\int r_1 \frac{(r_2-p_2)(r_2 p_1 - r_1 p_2)}{(\vec{r}-\vec{p})^2} e^{-ut(1-u_1)\vec{r}^2}\;,\quad J_1 \equiv (-2)\int p_1 \frac{(p_2-r_2)(p_2 r_1 - p_1 r_2)}{(\vec{p}-\vec{r})^2} e^{-ut(1-u_1)\vec{r}^2} \tag{20}$$

and

$$J_2 \equiv (-2)^2 \int \frac{(r_2-p_2)(r_2p_1-r_1p_2)}{(\vec{r}-\vec{p})^2} e^{-ut(1-u_1)\vec{r}^2} \frac{(p_2-r_2)(p_2r_1-p_1r_2)}{(\vec{p}-\vec{r})^2} \quad (21)$$

The integrals in (20) yield

$$J_0 + J_1 = (-2)\frac{\pi}{2}\int_0^\infty d\alpha \frac{e^{-\beta\vec{p}^2}}{(\alpha+z)^3}\{\alpha(p_1^2 - p_2^2) + u(p_1^2 + p_2^2)\} \quad (22)$$

with $= \alpha + z$, $z = ut(1 - u_1)$ and $\beta = \frac{\alpha z}{\alpha+z}$. Likewise one gets from (21)

$$J_2 = (-2)^2 \frac{\pi}{2}\int_0^\infty \alpha d\alpha \frac{e^{-\beta\vec{p}^2}}{u^4}\{\vec{p}^2 z^2 p_2^2 + \frac{1}{2}u(3p_1^2 + p_2^2)\} \quad (23)$$

On doing the integration one gets

$$J_0 + J_1 + J_2 = \frac{\pi}{a}(p_1^2 - p_2^2)e^{-z\vec{p}^2}\left\{(1 + e^{z\vec{p}^2}) + \frac{2z}{a}(1 - e^{z\vec{p}^2})\right\} \quad (24)$$

with $a = z^2\vec{p}^2$. Note that eq.(22) is antisymmetric to the exchange of $p_1$ and $p_2$ and will be further multiplied by $e^{-c\vec{p}^2}$ – where $c = t(1 - u(1 - u_1))$ as seen from eq.(17). This will alter (24) to

$$J_0 + J_1 + J_2 = \frac{\pi}{a}(p_1^2 - p_2^2)e^{-t\vec{p}^2}\left\{(1 + e^{z\vec{p}^2}) + \frac{2z}{a}(1 - e^{z\vec{p}^2})\right\} \quad (24a)$$

To this order therefore,

$$e^{-tC_1} = e^{-tp^2} + (-t)^2(-\frac{\lambda}{4\pi})^2\pi(p_1^2 - p_2^2)e^{-t\vec{p}^2}\int_0^1 udu \int_0^1 du_1 \frac{1}{a}\left\{(1 + e^{z\vec{p}^2}) + \frac{2z}{a}(1 - e^{z\vec{p}^2})\right\}$$

$$= e^{-t\vec{p}^2}\left\{1 + (\frac{\lambda t}{4\pi})^2 \pi(p_1^2 - p_2^2)\int_0^1 udu \int_0^1 du_1 \frac{1}{a}\left[(1 + e^{z\vec{p}^2}) + \frac{2z}{a}(1 - e^{z\vec{p}^2})\right]\right\} \quad (25)$$

with $a = z^2\vec{p}^2$, $z = ut(1 - u_1)$.

**The third order term**: Following (17) it can be written as

$$(-t)^3 \int_0^1 u^2 du \int_0^1 u_1 du_1 \int_0^1 du_2 e^{-t(1-u)p^2} \int \langle p|H_I|r\rangle e^{-xr^2}\langle r|H_I|q\rangle e^{-zq^2}\langle q|H_I|p\rangle e^{-tuu_1u_2\vec{p}^2} \quad (26)$$

with the integration now understood as $\int d^2r d^2q$, $x \equiv tu(1 - u_1)$ and $z \equiv tuu_1(1 - u_2)$. Using eqs.(18) and (19) the product of the matrix elements now becomes

$$\left(-\frac{\lambda}{4\pi}\right)^3 \left\{r_1 - 2\frac{(p_2-r_2)(p_2r_1-p_1r_2)}{(\vec{p}-\vec{r})^2}\right\}\left\{q_1 - 2\frac{(r_2-q_2)(r_2q_1-r_1q_2)}{(\vec{r}-\vec{q})^2}\right\}\left\{p_1 - 2\frac{(q_2-p_2)(q_2p_1-q_1p_2)}{(\vec{q}-\vec{p})^2}\right\} e^{-xr^2-zq^2}$$

$$(27)$$

of which only the following four will be apparently non –zero.

$$K_0 = \left(-\frac{\lambda}{4\pi}\right)^3 (-2)^2 r_1 \frac{(r_2-q_2)(r_2q_1-r_1q_2)}{(\vec{r}-\vec{q})^2} e^{-xr^2} \frac{(q_2-p_2)(q_2p_1-q_1p_2)}{(\vec{q}-\vec{p})^2} e^{-zq^2} \qquad (27)$$

$$K_1 = \left(-\frac{\lambda}{4\pi}\right)^3 (-2)^2 q_1 \frac{(p_2-r_2)(p_2r_1-p_1r_2)}{(\vec{p}-\vec{r})^2} e^{-xr^2} \frac{(q_2-p_2)(q_2p_1-q_1p_2)}{(\vec{q}-\vec{p})^2} e^{-zq^2} \qquad (28)$$

$$K_2 = \left(-\frac{\lambda}{4\pi}\right)^3 (-2)^2 p_1 \frac{(p_2-r_2)(p_2r_1-p_1r_2)}{(\vec{p}-\vec{r})^2} e^{-xr^2} \frac{(r_2-q_2)(r_2q_1-r_1q_2)}{(\vec{r}-\vec{q})^2} e^{-zq^2} \qquad (29)$$

$$K_3 = \left(-\frac{\lambda}{4\pi}\right)^3 (-2)^3 \frac{(p_2-r_2)(p_2r_1-p_1r_2)}{(\vec{p}-\vec{r})^2} \frac{(r_2-q_2)(r_2q_1-r_1q_2)}{(\vec{r}-\vec{q})^2} e^{-xr^2} \frac{(q_2-p_2)(q_2p_1-q_1p_2)}{(\vec{q}-\vec{p})^2} e^{-zq^2} \qquad (30)$$

We begin with $K_0$. On integrating over $\vec{r}$ one gets

$$K_0 = \left(-\frac{\lambda}{4\pi}\right)^3 (-2)^2 \frac{\pi}{2} \int_0^\infty d\alpha \frac{e^{-\beta \bar{q}^2}}{(\alpha+x)^3} \{\alpha(q_1^2-q_2^2)+(\alpha+x)q_2^2\} \frac{(q_2-p_2)(q_2p_1-q_1p_2)}{(\vec{q}-\vec{p})^2} e^{-zq^2}$$

$$= \left(-\frac{\lambda}{4\pi}\right)^3 (-2)^2 \left(\frac{\pi}{2}\right)^2 \int_0^\infty \int_0^\infty d\mu\, d\alpha \frac{e^{-(\beta+z)\bar{p}^2}}{2(\alpha+x)^3} \frac{e^{\sigma \bar{p}^2}}{(\beta+\mu+z)^4} \{xT_1+\alpha T_2\}$$

with $T_1 = p_1\{6\mu^2 p_2^2 + w(3-4\mu p_2^2)\}$, $T_2 = p_1\{2\mu^2(p_1^2-2p_2^2)+w(1+2\mu p_2^2)\}$, $w = \beta+\mu+z$ and $\sigma = \frac{(\beta+z)^2}{w}$, $\beta \equiv \frac{\alpha x}{\alpha+x}$. The integration over $\mu$ yields,

$$K_0 = \left(-\frac{\lambda}{4\pi}\right)^3 (-2)^2 \left(\frac{\pi}{2}\right)^2 \int_0^\infty d\alpha \frac{e^{-(\beta+z)\bar{p}^2}}{2(\alpha+x)^3} p_1\{xG_1+\alpha G_2\}$$

With $G_1 = -\left\{\frac{1}{a} 2p_2^2 + 3(p_1^2-3p_2^2)\frac{1}{a^2 \bar{p}^2}(-1+e^{a/b})+\left(\frac{b}{a^2}\right)\left(8p_2^2+(3p_1^2-p_2^2)e^{a/b}\right)\right\}$

$$G_2 = -\left\{\frac{1}{a} 2p_1^2 + 3(p_1^2-3p_2^2)\frac{1}{a^2 \bar{p}^2}(-1+e^{a/b})+\left(\frac{b}{a^2}\right)\left(8p_2^2+(3p_1^2-p_2^2)e^{a/b}\right)\right\}$$

and $a = (\beta+z)^2 \vec{p}^2$. Similarly,

$$K_1 = \left(-\frac{\lambda}{4\pi}\right)^3 (-2)^2 \frac{\pi}{2} \int_0^\infty d\alpha \frac{e^{-\beta \bar{p}^2}}{(\alpha+x)^2} p_1 q_1 \frac{(q_2-p_2)(q_2p_1-q_1p_2)}{(\vec{q}-\vec{p})^2} e^{-zq^2}$$

$$= \left(-\frac{\lambda}{4\pi}\right)^3 (-2)^2 \left(\frac{\pi}{2}\right)^2 p_1 e^{-(x+z)\vec{p}^2} \frac{1}{b}(-1+e^{x\vec{p}^2})\left\{\left(-\frac{1}{c}\right)(p_1^2-p_2^2 e^{z\vec{p}^2})+\frac{1}{c^2}z(p_1^2-p_2^2)(1-e^{z\vec{p}^2})\right\} \qquad (32)$$

with $b = x^2 \vec{p}^2$ and $c = z^2 \vec{p}^2$.

Likewise, with $\beta \equiv \frac{\alpha z}{\alpha+z}$

$$K_2 = \left(-\frac{\lambda}{4\pi}\right)^3 (-2)^2 \frac{\pi}{2} \int_0^\infty d\alpha \frac{e^{-(\beta+x)\vec{r}^2}}{(\alpha+z)^2} p_1 r_1 \frac{(p_2-r_2)(p_2r_1-p_1r_2)}{(\vec{p}-\vec{r})^2}$$

$$= \left(-\frac{\lambda}{4\pi}\right)^3 (-2)^2 \left(\frac{\pi}{2}\right)^2 \frac{p_1}{z^2}(S_1+S_2)$$

With

$$S_1 = \int_0^\infty d\lambda \, e^{-b/w}\left\{-\frac{1}{b}(p_1^2 - p_2^2 e^{b/w}) - \frac{1}{b^2}w(p_1^2 - p_2^2)(1 - e^{b/w})\right\}, w = \lambda + x + z, b = w^2\vec{p}^2 \quad (33)$$

$$S_2 = \int_0^\infty d\lambda \, e^{-c/u}\left\{-\frac{1}{c}(p_1^2 - p_2^2 e^{c/u}) - \frac{1}{c^2}u(p_1^2 - p_2^2)(1 - e^{c/u})\right\}, u = \lambda + x, c = u^2\vec{p}^2 \quad (34)$$

Finally,

$$K_3 =$$

$$(-\frac{\lambda}{4\pi})^3(-2)^3 \frac{\pi}{2}\int_0^\infty \int_0^\infty d\alpha d\beta \frac{e^{\vec{w}^2/t}}{t^5}\{t^3 r_2 p_2 \vec{r}\cdot\vec{p} + t^2 B_2 + tB_1 - 8\alpha\beta w_2^2(r_1 p_2 - r_2 p_1)^2\}\frac{(p_2-r_2)(p_2 r_1 - p_1 r_2)}{(\vec{p}-\vec{r})^2}e^{-(x+\alpha)\vec{r}^2 - \beta\vec{p}^2}$$

$$(35)$$

with $\vec{w} = \alpha\vec{r} + \beta\vec{p}$ and $t = \alpha + \beta + z$ and

$$B_1 = r_2 p_2(w_1^2 + w_2^2) + 3w_2(p_2 r_1 - p_1 r_2)(\beta p_1 - \alpha r_1) + 2\alpha\beta w_2(r_2 + p_2)(p_2 r_1 - p_1 r_2)^2$$

$$B_2 = \frac{1}{2}(3r_1 p_1 + r_2 p_2) - 2\alpha\beta r_2 p_2(p_2 r_1 - p_1 r_2)^2 - w_2(r_2 + p_2)\vec{r}\cdot\vec{p} - (p_2 r_1 - p_1 r_2)(\beta p_1 - \alpha r_1)(r_2 + p_2)$$

Integrating (35) over α gives

$$K_3 = (-\frac{\lambda}{4\pi})^3(-2)^3 \frac{\pi}{2}\int_0^\infty d\beta \sum_{j=0}^9 C_j e^{-z\vec{r}^2 - \beta(\vec{r}-\vec{p})^2}\frac{(p_2 - r_2)(p_2 r_1 - p_1 r_2)}{(\vec{p}-\vec{r})^2}e^{-x\vec{r}^2}$$

with each of the $C_i$ given below.

We now take up the integration over $\vec{r}$ with $C_0$ as an example, it being given by

$$C_0 = \frac{1}{f}\left[\begin{array}{c} r_1(r_2 - p_2)a + 6\beta r_2^2 a^2 \\ +\frac{1}{k^2}\{z^2 r_2 p_2 \vec{r}\cdot\vec{p} + \frac{1}{2}k(r_2 p_2 + 3r_1 p_1) + \beta^2 p_1(p_2 - r_2)a + \beta z[r_2^2\vec{p}^2 - p_1 p_2 a - p_2^2 \vec{r}\cdot\vec{p}]\}e^{f/k} \end{array}\right]$$

with $f = [\beta\vec{p} - k\vec{r}]^2$, $k = \beta + z$ and $a = r_1 p_2 - r_2 p_1$ ;

defining $\qquad I_0 \equiv \int d^2r \, C_0 \, e^{-z\vec{r}^2 - \beta(\vec{r}-\vec{p})^2}\frac{(p_2-r_2)(p_2 r_1 - p_1 r_2)}{(\vec{p}-\vec{r})^2}e^{-x\vec{r}^2}$

we obtain

$$I_0 \equiv \frac{\pi}{2}\int_0^\infty d\mu \int_0^\infty \frac{dv}{l^5}e^{m^2\vec{p}^2/l}(-A_0 + A_1)e^{-(\beta(\mu\beta+1)+v)\vec{p}^2} + \frac{\pi}{2}\int_0^\infty d\mu \int_0^\infty \frac{dv}{b^4}e^{n^2\vec{p}^2/b}A_2 e^{-\{v+\beta(\mu\beta+\frac{z}{k})\}\vec{p}^2}$$

$$(36)$$

with $l = (\beta + z)(\mu(\beta + z) + 1) + v + x$, $m = \beta(\mu(\beta + z) + 1) + v$, $n = \beta\mu(\beta + z) + v$ and $b = \mu(\beta + z)^2 + v + x$,

$$A_0 = p_1\left[p_2^2\vec{p}^2 m^3 + l\left(\frac{3}{2}m(p_1^2 - p_2^2) - 2m^2 p_2^2 \vec{p}^2\right) + l^2(mp_2^2\vec{p}^2 + 2p_2^2)\right]$$

$$A_1 = \frac{9\beta}{l}p_1\left(-2m^4 p_2^6 + 3lm^2 p_2^2\vec{p}^2 + l^2\left(p_1^2 + \frac{3}{2}\vec{p}^2 - 2mp_2^2\vec{p}^2\right)\right)$$

$$A_2 = p_1 \frac{1}{k^2} \left[ z^2 \{2n^2 p_2^2 \vec{p}^2 + b p_2^2 [1 - n(2p_2^2 + p_1^2)]\} + bk \left( \frac{1}{2} n(3p_1^2 - p_2^2) + b p_2^2 \right) \right]$$

$$+ p_1 \frac{1}{k^2} \left[ \beta^2 \left\{ n^2 p_2^2 \vec{p}^2 + b \left[ \frac{1}{2} (3p_1^2 + p_2^2) - 2n p_2^2 \vec{p}^2 \right] + b^2 p_2^2 \vec{p}^2 \right\} + \beta z \vec{p}^2 \left\{ \frac{3}{b} n^2 p_2^2 + \left( 3 - \frac{1}{4} n p_2^2 \right) - b p_2^2 \right\} \right]$$

Similarly, with

$$C_1 = \frac{1}{f^2} \left\{ \left[ \frac{1}{2} (3r_1 p_1 + r_2 p_2) + \beta \{ p_2 \vec{r} \cdot \vec{p} (r_2 - p_2) - p_2 (p_1 + r_1) a - 2a^2 \} + 2\beta^2 a^2 p_2 (p_2 - 7r_2) \right] \right.$$

$$\left. + e^{f/k} \left[ -\frac{1}{2} (3r_1 p_1 + r_2 p_2) + \beta (-2a^2 - a p_2 r_1 + r_2 p_2 [\vec{p}^2 + 3\vec{r} \cdot \vec{p}]) + 2\beta^2 a^2 p_2 (p_2 + r_2) \right] \right\}$$

we define
$$I_1 \equiv \int d^2r \, C_1 \, e^{-z\vec{r}^2 - \beta(\vec{r} - \vec{p})^2} \frac{(p_2 - r_2)(p_2 r_1 - p_1 r_2)}{(\vec{p} - \vec{r})^2} e^{-x\vec{r}^2} \tag{37}$$

to get

$$I_1 =$$

$$\frac{\pi}{2} \int_0^\infty \mu \, d\mu \int_0^\infty dv \frac{e^{m^2 \vec{p}^2/l}}{l^3} p_1 \{B_0 + \beta B_1 + 2\beta^2 B_2\} e^{-(\beta(\mu\beta+1)+v)\vec{p}^2} + \frac{\pi}{2} \int_0^\infty \mu \, d\mu \int_0^\infty dv \frac{e^{n^2 \vec{p}^2/b}}{b^3} p_1 \{B_3\} e^{-\left\{v + \beta\left(\mu\beta + \frac{z}{k}\right)\right\}\vec{p}^2}$$

with

$$B_0 = \frac{1}{2} m(3p_1^2 - p_2^2) + l p_2^2$$

$$B_1 = \frac{1}{l} (3m^2 p_2^2 \vec{p}^2 - l[3p_1^2 - p_2^2 + 2m p_2^2 \vec{p}^2] - l^2 m^2 p_2^2 \vec{p}^2)$$

$$B_2 = 3 \frac{p_2^2}{l^2} (5m^3 p_2^2 p_1^2 - 7lm\vec{p}^2 + 4l^2 \vec{p}^2)$$

$$B_3 = -\left( \frac{1}{2} n(3p_1^2 - p_2^2) + b p_2^2 \right) + \beta \left\{ \frac{7}{b} n^2 p_2^2 \vec{p}^2 - [3p_1^2 + p_2^2 + 2n p_2^2 \vec{p}^2] - b^2 p_2^2 \vec{p}^2 \right\} + \frac{6}{b} \beta^2 n p_2^2 \vec{p}^2 \quad ,$$

Continuing , we use

$$C_2 =$$

$$\frac{k}{f^2} [\{a r_1 (3r_2 - p_2) + r_2^2 (p_2^2 - \vec{r} \cdot \vec{p}) + r_1 r_2 p_1 p_2 + \beta a^2 (20 r_2^2 - 2 r_2 p_2)\} + e^{f/k} \{(r_2 + p_2)(a r_1 - p_2 \vec{r} \cdot \vec{p}) - 2\beta a^2 r_2 p_2\}]$$

$$C_3 = \frac{\beta^2}{f^3} \{(r_2 p_2 \vec{p}^2 + 6 p_1 p_2 a)(-1 + e^{f/k}) + 16 \beta a^2 p_2^2 (1 + 2 e^{f/k})\}$$

$$C_4 = \frac{\beta k}{f^3} \{(4 r_2 p_2 \vec{r} \cdot \vec{p} - 6a^2 + 4\beta p_2 (r_2 + p_2) a^2)(1 - e^{f/k}) - 32 \beta r_2 p_2 a^2 (2 + e^{f/k})\}$$

$$C_5 = \frac{k^2}{f^3}\{(2r_2^2\vec{r}\cdot\vec{p} - 4r_1r_2 a + 48\beta a^2 r_2^2) - 4\beta r_2(p_2+r_2)a^2(-1+e^{f/k})\}$$

$$C_6 = -\beta^2 \frac{p_2}{f^2 k} e^{f/k}\{-4p_1 a + 2p_2\vec{r}\cdot\vec{p} + 8\beta a^2 p_2^2\}$$

$$C_7 = -48\beta \frac{k^3}{f^4} a^2 r_2^2(-1+e^{f/k})$$

$$C_8 = 48\frac{\beta^3 k}{f^4} a^2 p_2^2(1-e^{f/k})$$

$$C_9 = -96\beta^2 \frac{k^2}{f^4} r_2 p_2 a^2(1-e^{f/k})$$

to define for all $k \geq 2$

$$I_k \equiv \int d^2 r\, C_k\, e^{-z\vec{r}^2 - \beta(\vec{r}-\vec{p})^2 \frac{(p_2-r_2)(p_2 r_1 - p_1 r_2)}{(\vec{p}-\vec{r})^2}} e^{-x\vec{r}^2} \tag{38}$$

The calculation of the integrals involved in each of the terms above is tedious and will be reported elsewhere; simultaneously an effort will be made to present the result of the calculation in a manner so as to appear as an extension of eq.(25) to the third order in λ.

To conclude, we have used the Schwinger expansion [7] in this paper to rework the Antonsen-Bormann idea[5,6]to obtain the ζ(s) function to second order in the gravitational constant G for the Lagrangian density (1) in 2 + 1 dimensional curved space, the metric for the latter being defined by the stationary solutions[3,4] of the Einstein field equations.

### Acknowledgements

A preliminary version of this work was presented at FFP11 that was held in Paris, France from July 6 – 9, 2010 and this was improved upon at the presentation at QTS7 that was held at Prague from Aug.7 – 13, 2011.I thank the respective organizers for their generous invitation to the two conferences and am grateful to the Indian Institute of Technology Madras for their financial support. I thank K.P.Deepesh, Ravi Shankar and G.Krishna Kumar for their generous assistance in the preparation of this manuscript. While working on this paper I have had useful correspondence with D.Broadhurst of The Open University, Milton Keynes, U.K , S.Laporta of INFN, Bologna,Italy and Ivan Gonzalez of the Universidad Santa Maria , Valparaiso, Chile and thank them for it..